\documentclass[journal=jacsat,manuscript=article]{achemso}

\usepackage[version=3]{mhchem} 
\usepackage{chemfig}
\usepackage[final]{pdfpages}

\author{Ernst D. Larsson}
\affiliation{Department of Theoretical Chemistry, Lund University, Chemical Centre, P. O. Box 124, SE-221 00 Lund, Sweden}
\author{Geng Dong}
\affiliation{Department of Theoretical Chemistry, Lund University, Chemical Centre, P. O. Box 124, SE-221 00 Lund, Sweden}
\alsoaffiliation{Department of Biochemistry and Molecular Biology, Shantou University Medical College, Shantou 514041, Guangdong, PR China}
\author{Valera Veryazov}
\affiliation{Department of Theoretical Chemistry, Lund University, Chemical Centre, P. O. Box 124, SE-221 00 Lund, Sweden}
\author{Ulf Ryde}
\affiliation{Department of Theoretical Chemistry, Lund University, Chemical Centre, P. O. Box 124, SE-221 00 Lund, Sweden}
\author{Erik D. Hedeg\aa rd}
\email{erik.hedegard@teokem.lu.se}
\affiliation{Department of Theoretical Chemistry, Lund University, Chemical Centre, P. O. Box 124, SE-221 00 Lund, Sweden}

\title[]{Is Density Functional Theory Accurate for Lytic Polysaccharide Monooxygenase Enzymes?}

\abbreviations{IR,NMR,UV}

\begin{document}

\begin{tocentry}
\centering
\includegraphics[scale=0.32]{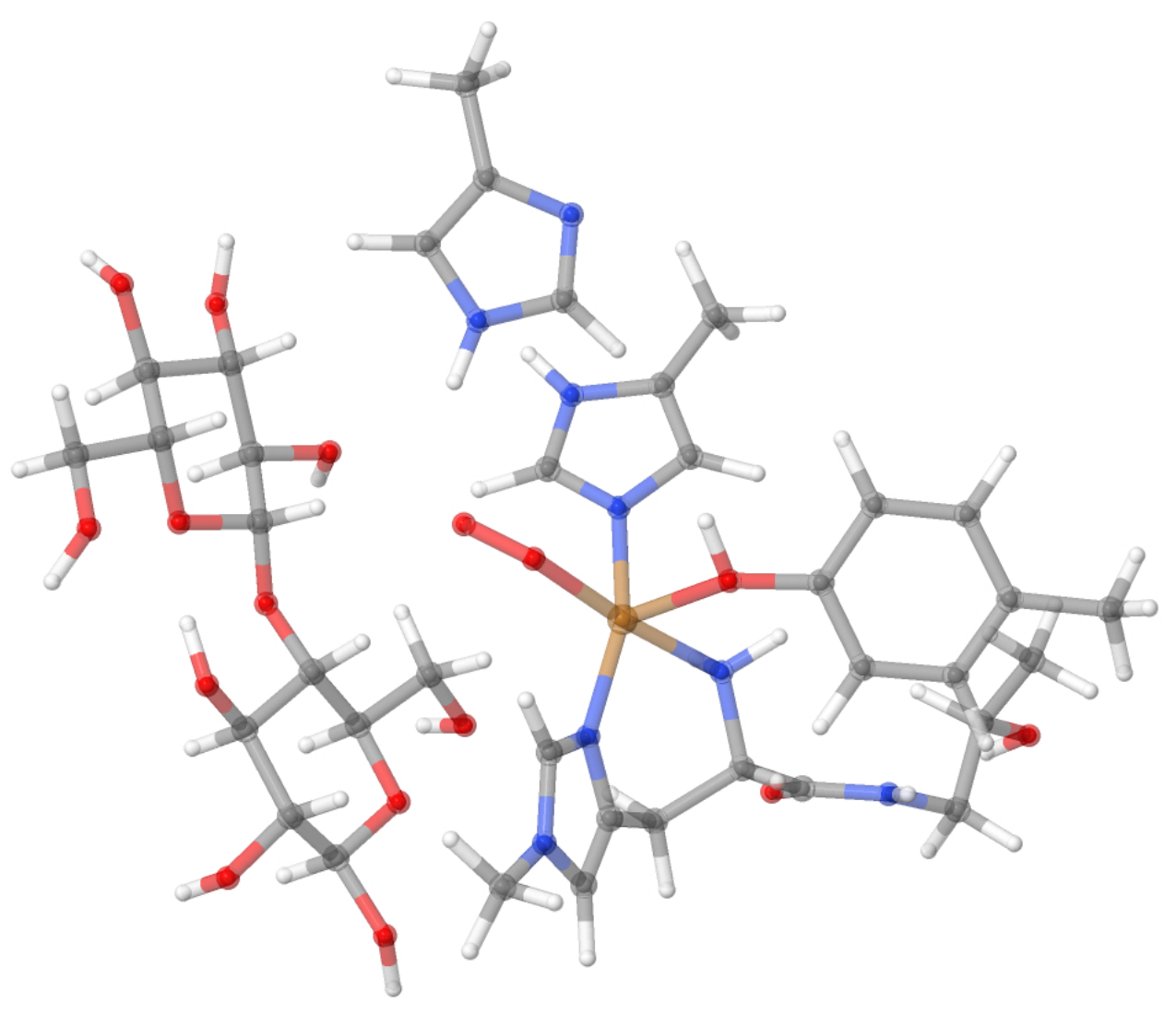}

\end{tocentry}

\begin{abstract}

The lytic polysaccharide monooxygenase (LPMO) enzymes boost polysaccharide depolymerization through oxidative chemistry, which has fueled the hope for more energy-efficient production of biofuel. We have recently proposed a mechanism for the oxidation of the polysaccharide substrate (Hedeg{\aa}rd \& Ryde, \textit{Chem. Sci.} 2018, \textbf{9}, 3866). In this mechanism, intermediates with superoxide, oxyl, as well as hydroxyl (i.e. \ce{[CuO2]+}, \ce{[CuO]+} and \ce{[CuOH]^{2+}}) cores were involved. These complexes can have both singlet and triplet spin states, and both spin-states may be important for how LPMOs function during catalytic turnover.   
Previous calculations on LPMOs have exclusively been based on density functional theory (DFT). However, different DFT functionals are known to  display large differences for  spin-state splittings in transition-metal complexes, and this has also been an issue  for LPMOs. In this paper, we study the accuracy of  DFT for spin-state splittings in  superoxide, oxyl, and  hydroxyl intermediates  involved in LPMO turnover. As reference we employ  multiconfigurational perturbation theory (CASPT2).   
\end{abstract}

\section{Introduction}

Atmospheric oxygen is believed to have been introduced in our atmosphere 2.0--2.5 billion years ago.\cite{kump2008} Nature has since devised numerous ways to exploit \ce{O2} to perform biochemical transformations, often by the use of transition metals. Copper is one of the metals employed for \ce{O2} activation by many enzyme families.\cite{solomon2014} A relatively new member of the \ce{O2}-activating enzymes is lytic polysaccharide monooxygenase (LPMO). The LPMOs were discovered in 2010\cite{harris2010,vaaje-kolstad2010} and were shown to boost polysaccharide depolymerization through oxidative chemistry. This was a paradigm shift in our understanding of how highly stable polysaccharides, such as  cellulose, are decomposed, which previously was believed to be solely hydrolytic. 
The discovery of LPMOs have fueled the hope for  production of biofuels from cellulosic biomass\cite{hemsworth2013a,beeson2015,span2015,walton2016,meier2018}, which could reduce the cost of biofuel production, because cellulose is cheap and  estimated to be the most abundant polysaccharide on Earth\cite{klemm2005}. 

\begin{figure}[htb!]
\centering
\includegraphics[width=0.90\textwidth]{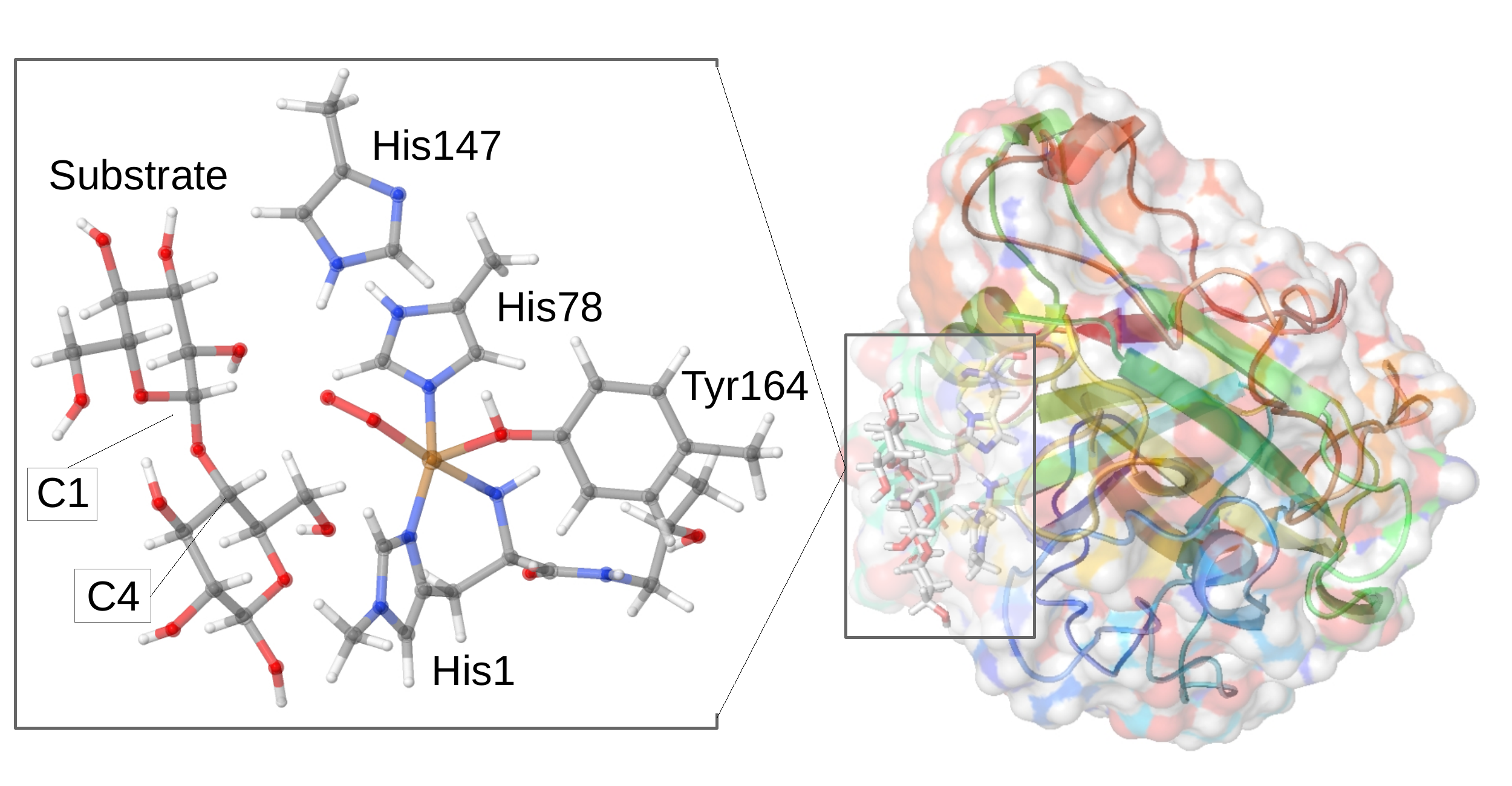}
\caption{General structure of AA9 LPMO \textit{Ls}(AA9)A, in complex with a substrate (taken from the 5ACF structure\cite{frandsen2016}), as well as the structure of the active site, from the QM/MM optimized structure of the \ce{[CuO2]+} intermediate\cite{hedegaard2018}, illustrating the employed QM system.} \label{activesite}
\end{figure}

The overall oxidation reaction of the LPMOs  involves \ce{O2} and two reduction steps (cf. Scheme \ref{netlpmoeaction}). 
\begin{scheme} 
 \schemestart
 \chemname{\ce{R-H}}{} +   \chemname{\ce{O2}}{} +  \chemname{\ce{2H+}}{} + 
  \chemname{\ce{2e-}}{}
\arrow(.base east--.base west){->[LPMO][][4pt]}  \chemname{\ce{R-OH}}{} + \chemname{\ce{H2O}}{} 
\schemestop 
 \caption{Reaction catalyzed by LPMO, where \ce{R-H} denotes a polysaccharide.\label{netlpmoeaction}  }
\end{scheme} 
However, it should be noted that this reaction may evolve through initial generation of peroxide as it was recently shown that both \ce{O2} and \ce{H2O2} can be employed as co-substrate.\cite{bissaro2017,hangasky2018} Further, the substrate-free LPMO can activate \ce{O2} and produce \ce{H2O2}\cite{kittl2012,kjaergaard2014,caldararu2019}.  The active site responsible for this chemistry is shown in  Figure \ref{activesite} in a form where \ce{O2} is bound to Cu together with a part of the substrate. The figure also shows the two substrate carbon positions (C1 and C4) oxidised by LPMOs; the positions are at the glycoside link between the sugar units. The active site itself is comprised of the copper ion, coordinated by two histidine residues, one of which coordinates with both the imidazole sidechain and the (terminal) amino group. This active site is conserved among all known LPMOs, which otherwise show a rather large sequence variation.\cite{tian2011,berka2011} This variation is evident even close to the active site. In the AA9 family of the LPMOs (which is the focus of the current article), there is a nearby tyrosine residue that can act as Cu ligand (depending on the copper oxidation state), whereas some other LPMO families lack this residue.

Regardless of whether \ce{O2} or \ce{H2O2} is used as co-substrate, the mechanism employed is still to a large degree unknown: most structural information is obtained for the LPMO resting state without substrate or co-substrates bound.\cite{quinlan2011,phillips2011,beeson2012,horn2012,forsberg2011,hemsworth2013b,vaaje-kolstad2012,aachmann2012,vaaje-kolstad2013,vaaje-kolstad2017}. A few crystal structures of \ce{O2}-bound LPMOs\cite{dell2017,bacik2017} or LPMOs complexed with substrates\cite{frandsen2016,simmons2017} have been reported. However, no structural data have been obtained for oxygen-bound intermediates complexed with substrates. Thus, the intermediate shown in Figure \ref{activesite} was obtained from a combined quantum mechanics and molecular mechanics (QM/MM) optimization.\cite{hedegaard2018} These calculations further showed that the species in Figure \ref{activesite} is best described as superoxide ion (\ce{O2$^{\cdot}$-}) bound to a Cu(II) ion. 

A key part of the LPMO mechanism is believed to be the abstraction of a hydrogen atom from the \ce{C-H} bond in the C1 or C4 positions of the polysaccharide substrate. Despite efforts from both theory and experiments, the active species that abstracts the hydrogen from the polysaccharide substrate has been the matter of controversy.
 Some suggestions for the mechanism employs a superoxide for the \ce{C-H} abstraction\cite{phillips2011,beeson2012,li2012,beeson2015}, but other suggestions have involved hydrogen abstraction from an  
oxyl (\ce{O$^{\cdot}$-}) species\cite{kim2014,beeson2015,lee2015,walton2016}. Studies on model systems have also suggested hydroxy\cite{dhar2015,walton2016} and hydroperoxy complexes\cite{neisen2017} as the reactive species. 

Rather few studies have addressed the mechanism of LPMOs with quantum mechanical (QM) methods.\cite{kim2014,kjaergaard2014,bertini2017,wang2018,hedegaard2017a,hedegaard2017b,hedegaard2018}. We and a few other groups have recently initiated investigations of the LPMO mechanism employing both QM-cluster and QM/MM calculations.\cite{hedegaard2017a,hedegaard2017b,hedegaard2018} From these calculations we could show that hydrogen abstraction by complexes involving an intact \ce{O-O} is not energetically feasible\cite{hedegaard2017b}, whereas both oxyl\cite{kim2014,hedegaard2017b,hedegaard2018} or hydroxy complexes\cite{hedegaard2017b,hedegaard2018} are more reactive. 

While important mechanistic insight thus have been obtained from computational studies, all investigations so far have relied exclusively on density functional theory (DFT). Yet, we have for some intermediates shown how the reaction as well as spin-state energetics depend quite strongly on the choice of the DFT functional.\cite{hedegaard2018} Moreover, in metalloproteins whose active sites resemble LPMOs\cite{Melia2013,Abad2014,chen2004,Crespo2006} 
 both Cu--superoxide and Cu--oxyl species are known to involve electronic structures where DFT occasionally fails to predict the correct ground state.\cite{Gagliardi2009} 

In this study, we address the performance of DFT for the spin-state splittings in several LPMO intermediates that have been shown to be important in various parts of the LPMO mechanism, employing  QM/MM optimised structures\cite{hedegaard2018} from our previous studies.  The target intermediates are the superoxide, oxyl and hydroxyl complexes, i.e., intermediates with \ce{[CuO2]+}, \ce{[CuO]+}, and \ce{[CuOH]^{2+}} cores, respectively. The first intermediate is included although it is probably not relevent for \ce{C-H} abstraction. However, it is still the only  structurally characterized intermediate\cite{dell2017,bacik2017} after introduction of \ce{O2}, and its inclusion  more firmly connects our study with experimentally observed intermediates.
All three intermediates  have singlet and triplet spin states close in  energy according to DFT predictions\cite{hedegaard2018} and thus both states may be involved in the mechanism. While we are not aware of any experimental studies directly probing the spin-state splitting of the investigated species, we have for the \ce{[CuO2]+} intermediate previously confirmed that a superoxide provides the best fit to the experimental structure\cite{caldararu2019} and this interpretation complies with close lying triplet  and singlet spin states.

The performance of DFT will here be estimated using  a reference method and for this purpose we employ multiconfigurational perturbation theory to the second order, based on a complete active-space wavefunction (CASPT2)\cite{andersson1990,andersson1992}, which we compare to the results of several popular DFT functionals.

\section{Computational Details}

We study three different intermediates. All structures were taken from our previous QM/MM calculations\cite{hedegaard2018} in which they were optimized in triplet and singlet spin-states, respectively. All were optimized with the same size of QM region; an example is provided in Figure \ref{activesite} for the intermediate with \ce{O2} bound to copper. We will here  use the short-hand notation  \ce{[CuO2]+}, \ce{[CuO]+} and \ce{[CuOH]^{2+}}, for these three intermediates.  
 The QM/MM calculations and their setup were described in more detail in Ref.~\citenum{hedegaard2018}. Here we only note that the setup is based on the  crystal  structure from Frandsen \textit{et al.}\cite{frandsen2016} (PDB: 5ACF) and the QM system included the imidazole ring of His78 and the phenol ring of Tyr164, both 
capped with a hydrogen atom replacing C$^{\alpha}$. The entire His1 residue, which coordinates to  Cu through both the terminal amino group and the imidazole sidechain, was also included and the neighboring Thr2 residue was included up to the C$^{\alpha}$ atom, which was replaced by a hydrogen atom.  The fifth ligand (\textit{trans} to the terminal \ce{NH2} group) was either superoxide, oxyl or hydroxide, giving rise to \ce{[CuO2]+}, \ce{[CuO]+} and \ce{[CuOH]^{2+}} species, respectively. 
In addition, the QM system also contained two glucose rings (from the substrate) and a second-sphere histidine residue (His147, protonated on the N$^{\epsilon 2}$ atom). 

The three intermediates have electronic structures that can attain both singlet and triplet spin states. When necessary, we designate the spin state of the intermediate by giving the spin multiplicity ($2S+1$) in superscript, e.g., $^3$\ce{[CuO2]+} and $^1$\ce{[CuO2]+} in case of the superoxide species. The singlet states $^1$\ce{[CuO2]+} and $^1$\ce{[CuO]+}  were of open-shell nature and were optimized as unrestricted open-shell (broken-symmetry) singlets. The open-shell nature was confirmed by analyzing the spin-densities (cf.~Table \ref{d-orbital-contributions}). The comparison between triplet and open-shell singlet were done directly from the obtained energies and no correction schemes were employed. For $^1$\ce{[CuOH]^{2+}}, attempts to optimize the open-shell singlet yielded essentially the closed-shell state (the resulting spin densities were small; the Cu and O  atoms in the $^1$\ce{[CuOH]^{2+}} moiety carried spin densities of 0.09 and 0.07. respectively). Energies of the closed- and open-shell states were within 1 kJ/mol and the structures were identical. Therefore, we used the structure and energies of the closed-shell singlet state.

\begin{figure}[htb!]
\centering
\includegraphics[width=1.00\textwidth]{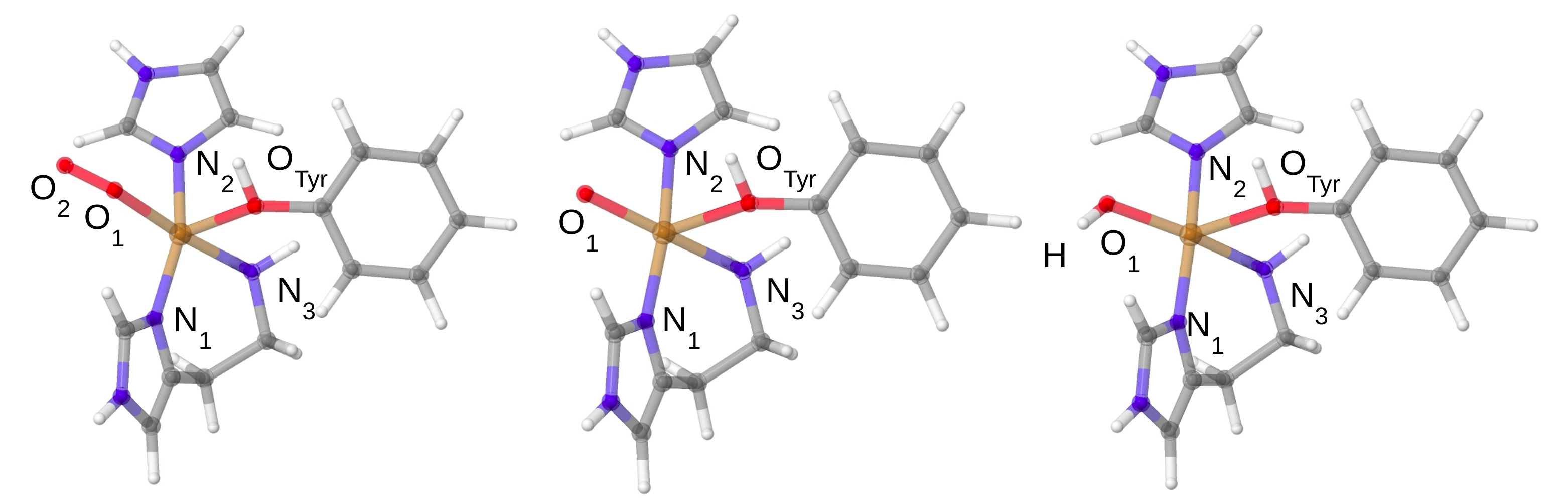}
\caption{Structures of the $^3$\ce{[CuO2]+}, $^3$\ce{[CuO]+} and $^3$\ce{[CuOH]^{2+}} complexes used in the CASPT2 calculations. The shown structures are all in triplet spin states because the structural differences between singlet and triplet are small, up to 0.04 {\AA} in the the first coordination sphere for \ce{[CuO2]+}  and \ce{[CuO]+} and up to 0.08 \AA\  for \ce{[CuOH]^{2+}} (cf. Table \ref{structure:tab}). \label{structure}} 
\end{figure}

Owing to the large computational cost, the CASPT2 calculations were performed on truncated systems (called model 1), compared to the QM system used in our QM/MM calculations\cite{hedegaard2018} (which are called model 2). The truncated model 1 for all three studied states are shown in Figure \ref{structure} and selected bond lengths are given in Table \ref{structure:tab}. As can be inferred from Figures \ref{activesite} and  \ref{structure}, the truncation in model 1 includes removal of the second-sphere histidine and the substrate, while methyl groups were replaced by hydrogen atoms. For the bidentate histidine ligand, the carbonyl C was replaced by a hydrogen atom (i.e. excluding all parts of Thr2). The structures were obtained by adding H atoms at a C--H distance of 1.10  \AA\ along the C--C bond (in the case of the one \ce{N-C} bond where C was replaced on His1, the \ce{N-H} distance is 1.01 \AA). No re-optimizations of the structures were performed in order to keep the structures as close to the protein structures as possible.
\begin{table}[htb!]
\centering
\caption{Bond distances (in {\AA}) and angles (in $^{\circ}$) for the considered complexes.  } 
\label{structure:tab}
\begin{tabular}{lccccccc}
\hline\hline
\\[-2.0ex]
Complex           & \ce{Cu-O1} & \ce{Cu-N1} & \ce{Cu-N2} & \ce{Cu-N3} &   \ce{Cu-O$_{\rm Tyr}$} & \ce{O1-O2}/\ce{H} & \ce{Cu-O1-O2}/\ce{H} \\       
 \hline \\[-2.0ex]
\ce{$^3$[CuO2]+}      & 2.08 & 1.96 & 1.98  & 2.11 & 2.29 & 1.28 & 116  \\[0.5ex]                    
\ce{$^1$[CuO2]+}      & 2.05 & 1.96 & 1.98  & 2.10 & 2.29 & 1.29 & 117  \\[0.5ex]
\ce{$^3$[CuO]+}       & 1.89 & 1.95 & 1.99  & 2.12 & 2.40 & -    & -    \\[0.5ex]   
\ce{$^1$[CuO]+}       & 1.85 & 1.95 & 1.95  & 2.11 & 2.40 & -    & -    \\[0.5ex]          
\ce{$^3$[CuOH]^{2+}}  & 1.93 & 1.96 & 2.00  & 2.13 & 2.32 & 0.99 & 123  \\[0.5ex]          
\ce{$^1$[CuOH]^{2+}}  & 1.85 & 1.90 & 1.95  & 2.05 & 2.35 & 0.99 & 117  \\[0.5ex]  
\hline \hline
\end{tabular}
\end{table} 

The CASSCF/CASPT2\cite{andersson1992} calculations were performed  with MOLCAS 8.2.\cite{Aquilante2016}. 
The  selected active space included 12 electrons in 12 orbitals, denoted CAS(12,12), for \ce{[CuO2]+} and CAS(14,16) for \ce{[CuO]+} and \ce{[CuOH]^{2+}}.  The selection of orbitals for the active space for the various species is described in more detail in the Results section. We used two different basis sets in the CASSCF/CASPT2 calculations. The first involved ANO-RCC-VQZP ([7s6p4d3f2g1h]) for Cu\cite{roos2005} and ANO-RCC-VTZP for the C, N and O ([4s3p2d1f]) for all) and [3s1p] for H. \cite{widmark1990} The second was a Dunning correlation-consistent (cc) basis set with cc-pwCVQZ-DK for Cu\cite{Balabanov2005}, cc-pVTZ-DK for C, N and O, and cc-pVDZ-DK for H.\cite{Dunning1989} Scalar relativistic effects were included with the Douglas--Kroll--Hess (DKH) approach to the second order.\cite{hess1986,reiher2004a,reiher2004b} For the CASPT2 calculations, all valence electrons and $3s3p$ semi-core electrons of copper were correlated. The evaluation of two-electron integrals in MOLCAS were approximated with Cholesky decomposition and using an on-the-fly generated auxiliary basis set \cite{RICD:Pedersen2009}. All PT2 calculations were performed with the standard ionisation potential--electron affinity (IPEA) Hamiltonian shift of 0.25 a.u.\cite{ghigo2004} 

The DFT calculations were done with the def2-TZVPP basis set on all atoms.\cite{weigend2005}  
We employed four different functionals, namely TPSS\cite{tao2003}, TPSSh\cite{staroverov2003}, B3LYP\cite{becke1993,becke1988,lee1988} and M06L.\cite{zhao2006} These functionals were chosen because they represent four commonly employed functionals employing different design strategies: the TPSS and M06L functionals are both meta-GGA but while the former was constructed to satisfy exact physical constraints without empirical parameters, the latter is heavily parameterized (and includes transition metal compounds in its parameterization). The TPSSh functional is a hybrid formulation of TPSS, adding 10 \%  exact Hartree-Fock exchange, while the B3LYP functional is an empirical hybrid functional that has been a standard method in quantum chemistry since its development in the early 90's. For non-hybrid functionals  the  DFT calculations were sped up by expanding the Coulomb interactions in an auxiliary basis set (the resolution-of-identity approximation)\cite{eichkorn1995,eichkorn1997}, employing standard def2-TZVPP auxiliary basis sets. The empirical D3 dispersion corrections were included with the Becke--Johnson damping.\cite{grimme2011} All DFT calculations were performed with the Turbomole 7.1 software.\cite{turbomole} 

We additionally carried out a series of coupled-cluster (CC) calculations employing CC with singles,  doubles and perturbative triples, CCSD(T) as implemented in Turbomole 7.1. The calculations employed the cc-pVTZ basis set\cite{Balabanov2005} for Cu and def2-SV(P)\cite{Schafer1992} for the other atoms. All the CCSD(T) calculations were spin restricted. However, the D1 amplitudes for the singlet states turned out to be rather large and in all cases over the threshold D1 $>$  0.15\cite{Jiang2012}, suggesting multiconfigurational character (0.24 for [\ce{CuO2]+}, 0.19 for \ce{[CuO]+} and 0.27 for \ce{[CuOH]^{2+}}). As discussed below, the CASPT2 calculations also indicated that the singlet states are multiconfigurational and hence we will not discuss the CCSD(T) results in detail.  

We finally note that on several occasions only values for one representative basis set or functional are reported. Thus, all shown orbitals are obtained with ANO type basis set. The Dunning-type basis set gave rise to identical orbitals. Similarly, the reported Mulliken spin densities were for all CASPT2 calculations obtained with ANO-type basis set and also here the Dunning basis set led to very similar results. With DFT, the reported Mulliken spin densities are obtained with the TPSS functional (and def2-TZVPP basis set) and also here the same conclusions are obtained with other functionals.

\section{Results} \label{results}

In this study, we compare energy splittings between the singlet and triplet spin states ($\Delta E_{\rm ts}$ = $E_{\rm triplet} - E_{\rm singlet}$) obtained with either DFT or CASPT2. We consider three LPMO intermediates, \ce{[CuO2]+}, \ce{[CuO]+} and \ce{[CuOH]^{2+}} and the results  are discussed in separate sections. 

\subsection{The superoxide state}
\begin{figure}
\centering
\includegraphics[scale=.8]{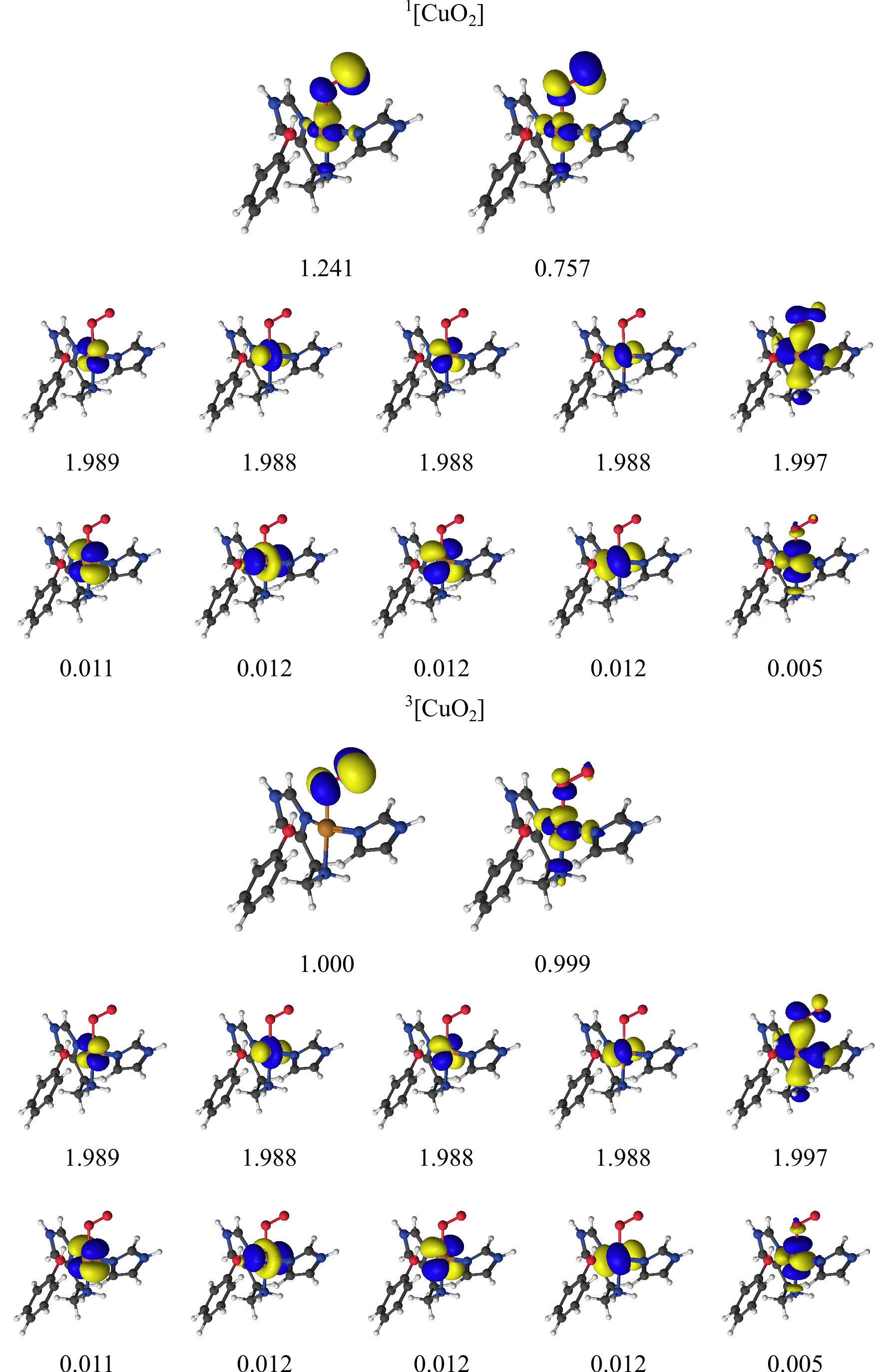}
\caption{Active orbitals and their occupation numbers from the CAS(12,12) calculation with  ANO-RCC type basis sets for the \ce{[CuO2]+} species.  } \label{cu-o2-edl-figure}
\end{figure}

Earlier studies on LPMO\cite{kim2014,kjaergaard2014,bertini2017,wang2018,hedegaard2017a,hedegaard2017b,hedegaard2018} and other copper enzymes\cite{Gagliardi2009,solomon2014} have shown that \ce{[CuO2]+} is often best interpreted as a doublet superoxide radical (\ce{O2-}) bound to a doublet Cu(II) ion. In DFT, this can give rise to either a triplet state or an open-shell (broken-symmetry) singlet state, depending on the alignment of the two unpaired spins. To describe the \ce{[CuO2]+} moiety, 
we employed a CAS(12,12) active space for both the singlet and triplet spin states. The chosen active-space orbitals are shown in Figure \ref{cu-o2-edl-figure} in combination with natural occupation numbers (shown below the orbitals).

\begin{table}[htb!]
\centering
\caption{The combined weight of copper $3d$ and oxygen ligand (from superoxide, oxyl or hydroxyl) in orbitals with occupation numbers significantly different from 2 or 0 in Figures \ref{cu-o2-edl-figure}, \ref{cu-o-edl-figure} and \ref{cu-oh-edl-figure}. The results are obtained with ANO basis sets.} 
\label{d-orbital-contributions} 
\begin{tabular}{lrrrrr}
\hline\hline    \\[-2.0ex]
Intermediate         &  Occ.    & $3d$ & $2p$   \\[0.5ex]
 \hline \\[-2.0ex]
$^3$\ce{[CuO2]+}     &  1.000   & 0.00 & 1.30   \\[0.5ex] 
$^3$\ce{[CuO2]+}     &  0.999   & 0.92 & 0.05   \\[0.5ex] 
$^1$\ce{[CuO2]+}     &  1.241   & 0.46 & 0.60   \\[0.5ex] 
$^1$\ce{[CuO2]+}     &  0.757   & 0.44 & 0.72   \\[0.5ex]
$^3$\ce{[CuO]+}      &  1.001   & 0.00 & 0.99   \\[0.5ex]
$^3$\ce{[CuO]+}      &  1.009   & 0.90 & 0.13   \\[0.5ex]
$^1$\ce{[CuO]+}      &  1.323   & 0.49 & 0.50   \\[0.5ex]
$^1$\ce{[CuO]+}      &  0.677   & 0.45 & 0.55   \\[0.5ex]
$^3$\ce{[CuOH]^{2+}} &  1.003   & 0.01 & 0.96   \\[0.5ex] 
$^3$\ce{[CuOH]^{2+}} &  1.005   & 0.95 & 0.05   \\[0.5ex] 
$^1$\ce{[CuOH]^{2+}} &  1.688   & 0.43 & 0.44   \\[0.5ex] 
$^1$\ce{[CuOH]^{2+}} &  0.310   & 0.53 & 0.46       \\[0.5ex]
\hline \hline
\end{tabular}
\end{table}
 The active space includes one bonding ligand orbital that is located between Cu, the three nitrogen atoms of the two histidine ligands and \ce{O2-}. This orbital has an occupation number close to two (1.997) for both states. It could be interpreted as \ce{O2}-antibonding orbital, albeit with large amplitude on histidine. We decided to include this orbital due to its large amplitude on \ce{O2-} in both singlet and triplet species. The active space is further comprised of the five $3d$ orbitals, of which four are doubly occupied with occupation numbers between 1.988 and 1.989 in both the singlet and triplet states. The fifth Cu $3d_{z^2}$ orbital interacts with the oxygen $\pi^\star$ orbitals and they show a pair of partly occupied orbitals: in the triplet state, they both have occupation numbers around 1.0 and are of rather pure $3d$ and oxygen $\pi^{*}$ character, respectively (see Table \ref{d-orbital-contributions} where the combined weight of the constituting atomic orbitals are given). With four of the five $3d$ orbitals doubly occupied, and the last $3d$ orbital singly occupied, an Cu(II) interpretation seems reasonable. In the singlet state, the singly occupied orbitals have occupation numbers of 1.241 and 0.757, respectively, and are more mixed between Cu  $3d$ and O $2p$ character (see Table \ref{d-orbital-contributions}).
 The remaining five orbitals have low occupation numbers (0.005--0.012) in both triplet and singlet spin states. They are included as a second shell of $d$ orbitals, which has previously shown to be important to obtain accurate CASPT2 energies.\cite{varyazov2011} Overall, the orbitals for the singlet and triplet states are similar, which is desired for ensuring accurate spin-state splittings. We also note that a similar active space was employed in the study of a protein with an active site resembling the one studies here.\cite{Gagliardi2009} 
 
For the singlet state, a closer investigation of the underlying CASSCF wavefunction reveals that two configurations have large weights, 0.61 and 0.37, respectively. The four $3d$ orbitals and the \ce{O2^{-}} orbital (with occupation number 1.997) are doubly occupied in both configurations, and they differ instead in the occupation of the two orbitals with occupation numbers 1.241 and 0.757. In the first configuration, the former is doubly occupied and the latter unoccupied, whereas the opposite is true for the second configuration; in combination with the fact that the two partially occupied orbitals are a mixture of copper $3d$ orbitals and oxygen $2p$ orbitals (cf.~table \ref{d-orbital-contributions}), we assign the oxidation state as Cu(II). Thus, both spin states represent primarily a Cu(II)--superoxide state, but with different spin couplings.  

The spin-state splittings from the CAS(12,12)PT2 calculations as well as the four investigated DFT functionals are shown in Table \ref{splittings}.  
The singlet--triplet splitting energies are quite small for all methods. The DFT methods predict the triplet state to be most stable, by 16--19 kJ/mol for TPSS, TPSSh and B3LYP and by 31 kJ/mol for M06-L. Interestingly, CASPT2 predicts that the singlet is 2 kJ/mol more stable than triplet state, meaning that the two states are essentially degenerate. 

The results in Table \ref{splittings} were obtained with model 1. For the B3LYP and TPSS functionals, we also estimated $\Delta E_{\rm ts}$ with the larger model 2. The results from this investigation are shown in Table \ref{splittings_model2}., showing that the larger model decreases the stability of the triplet state by 3--6 kJ/mol.
 
\begin{table}[htb!]
\centering
\caption{Singlet--triplet splitting energies $\Delta E_{\rm ts} = E_{ \rm t} - E_{\rm s}$ (kJ/mol)  obtained with different methods using model 1. The CASPT2 results are with ANO basis sets (results with Dunning-type basis sets shown in parentheses). All DFT results are obtained with def2-TZVPP basis set.} 
\label{splittings} 
\begin{tabular}{lrrrrr}
\hline\hline
\\[-2.0ex]
 \multicolumn{6}{c}{$\Delta E_{\rm ts}$ (kJ/mol)}  \\ [0.5ex]      
 \hline \\[-2.0ex]
Intermediate               &  CASPT2           & TPSS    & TPSSh  & B3LYP        & M06-L  \\[0.5ex]
 \hline \\[-2.0ex]
\ce{[CuO2]+}     &  2.4  (2.1)       & $-$16.0 &  $-$15.5  & $-$19.0  & $-$30.9 \\[0.5ex] 
 \ce{[CuO]+}     & $-$20.7 ($-$21.2) & 7.4     & 5.5      & 4.4      & $-$1.3 \\[0.5ex]
\ce{[CuOH]^{2+}} & 93.2  (92.2)      & 62.0      &40.0       & 30.7     & 49.3 \\[0.5ex] 
\hline \hline
\end{tabular}
\end{table}

Thus, our results show that care must be exercised when assessing relative spin-states with DFT for LPMO intermediates and that for the TPSS, TPSSh and B3LYP functionals, spin-states separated by 20 kJ/mol or less can essentially not be distinguished based on DFT. The M06-L functional is further off (by 30 kJ), although this functional was parameterized employing transition-metal systems.\cite{zhao2006} Interestingly, a previous theoretical study of a \ce{[CuO2]+} moiety with several different ligands\cite{Gagliardi2009} showed that CAS(12,12)PT2 predicts the triplet to be most stable for most ligands, but in one case, the singlet was found to be more stable. The energy differences varied between  13 kJ/mol and $-25$ kJ/mol, showing that the ligands can have large influence on the singlet--triplet splitting. Moreover, DFT does not always  predict this effect correct: in particular, the M06-L functional gave qualitatively wrong spin-state splittings on several occasions.\cite{Gagliardi2009}
\begin{table}[htb!]
\centering
\caption{Singlet--triplet splitting energies $\Delta E_{\rm ts} = E_{ \rm t} - E_{\rm s}$ (kJ/mol) obtained with different functionals and def2-TZVPP basis sets with model 2.} 
\label{splittings_model2} 
\begin{tabular}{lrrrr}
\hline\hline
\\[-2.0ex]
  & \multicolumn{4}{c}{$\Delta E_{\rm ts}$ (kJ/mol)}   \\       
 \hline \\[-2.0ex]
Intermediate & \multicolumn{2}{c}{TPSS} & \multicolumn{2}{c}{B3LYP} \\[0.5ex]  
& QM$^a$ & QM/MM$^b$ & QM$^a$ & QM/MM$^b$ \\
 \hline \\[-2.0ex]
\ce{[CuO2]+}      & --12.8 & --12.3 & --13.4 & --15.3 \\[0.5ex] 
\ce{[CuO]+}       &  --5.9 &  --6.9 &  --3.7 &  --3.5 \\[0.5ex]
\ce{[CuOH]^{2+}}  &   36.1 &   45.1 &   21.0 &   33.8 \\[0.5ex] 
\hline \hline
\end{tabular}
\\[0.5ex]
$^a$ Energy obtained in vacuum for the QM/MM optimised structures with model 2. \\ 
$^b$ The corresponding QM/MM energies (from Ref.~\citenum{hedegaard2018}).
\end{table}

\subsection{The oxyl state}

We next turn to the \ce{[CuO]+} species, which is among the states that we and others previously have suggested to be active in the \ce{C-H} abstraction.\cite{kim2014,hedegaard2017b,hedegaard2018}
The \ce{[CuO]+} intermediate has in previous studies (see e.g.~Ref.~\citenum{hedegaard2018}) been interpreted as a doublet \ce{O-} radical bound to doublet Cu(II), giving either a triplet or an open-shell singlet state.

\begin{figure}
\centering
\includegraphics[scale=.8]{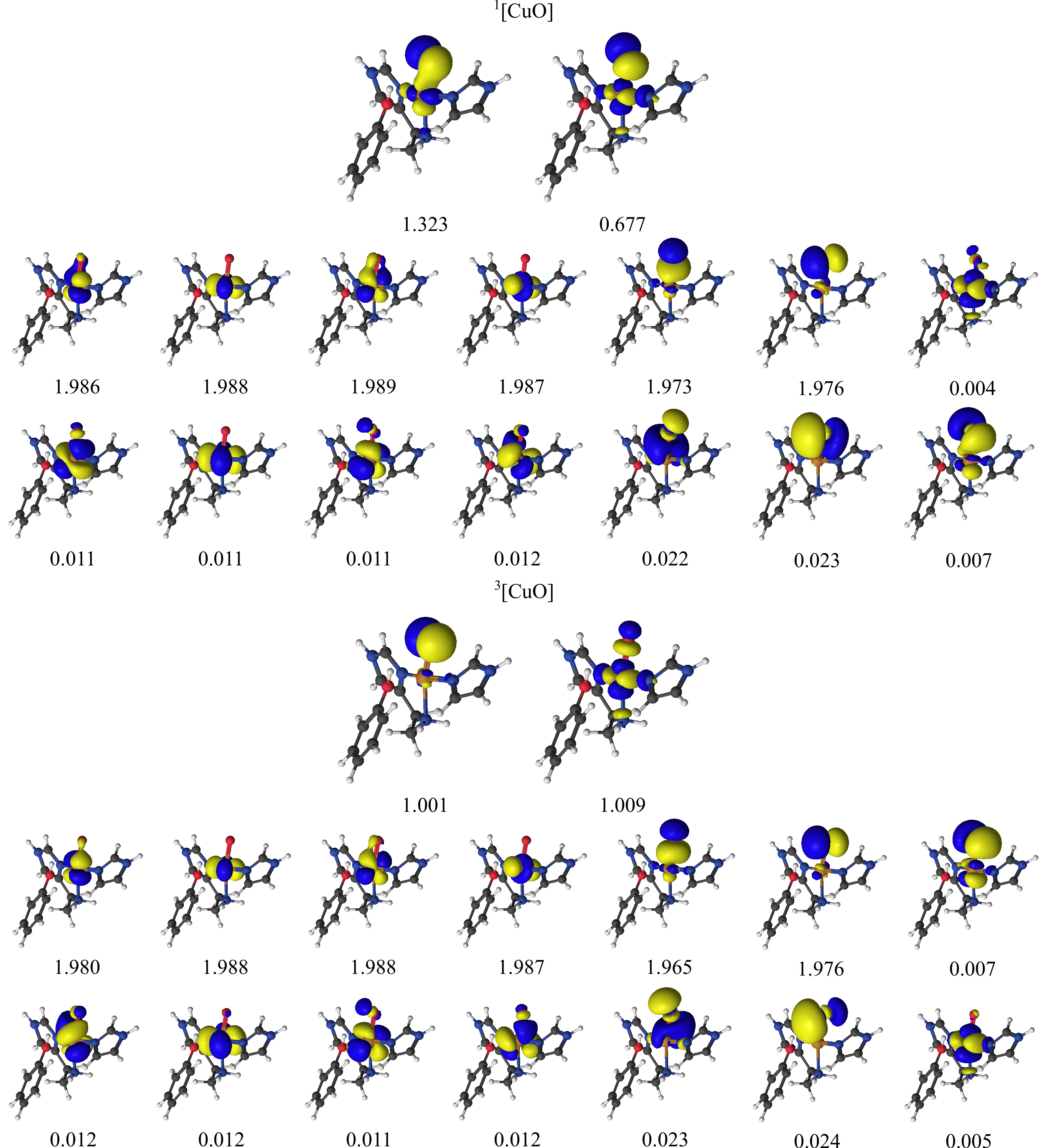}
\caption{ Active natural orbitals and their occupation numbers from a CAS(14,16) calculation with  ANO-RCC type basis sets for the \ce{[CuO]+} species.} \label{cu-o-edl-figure}
\end{figure}

The CASSCF active-space orbitals and the corresponding occupation numbers  are shown in Figure \ref{cu-o-edl-figure}. Our selected active spaces include the three oxygen $2p$ orbitals, the five Cu $3d$ orbitals and the five Cu $4d$ (double-shell) orbitals. In addition, we included the three O $3p$ orbitals, leading to the shown CAS(14,16) active space. This active space is slightly larger than the active space employed in Ref.~\citenum{Gagliardi2009}, due to the addition of the oxygen $3p$ orbitals.  As for the \ce{[CuO2]+} intermediate, the triplet state has two orbitals with occupation numbers close to 1. In \ce{[CuO]+}, these orbitals are of Cu $3d_{z^2}$ and O $2p$ character, showing that triplet \ce{[CuO]+}  can be interpreted as oxyl, i.e., with one unpaired electron on Cu(II) and another on \ce{O-}. The situation in the singlet is a bit less clear. It is again the orbitals of Cu $3d_{z^2}$ and O $2p$ character that have occupation numbers that deviate significantly from 0 or 2: the corresponding occupation numbers are 1.323 and 0.677, respectively (cf.~Figure \ref{cu-o-edl-figure}). Compared to the triplet, these two orbitals are  a mix of oxygen $2p$ and $3d$ orbitals, as seen for the \ce{[CuO2]+} intermediate (cf.~Table \ref{d-orbital-contributions}). Investigation of the underlying CASSCF wavefunction again shows some degree of multiconfigurational nature with two configurations having significant contributions to the total wavefunction with weights 0.62 and 0.32, respectively. Similarly to the \ce{[CuO2]+} state, both configurations have four doubly occupied Cu $3d$ orbitals as well as two doubly occupied ligand orbitals, in this case the oxyl $2p$ orbitals with occupation numbers 1.973 and 1.976. In addition, the configuration with the largest weight has the orbital with occupation number 1.323 occupied, whereas the orbital with occupation number 0.677 is  empty. The opposite is true for the second-largest configuration. With a similar argument as for the \ce{[CuO2]+} state, the large mixing of copper $3d$ orbital and ligand (oxygen) $2p$ orbitals for these two orbitals leads to an interpretation where Cu(II) and a \ce{O-} radical are spin-coupled to a singlet.

The singlet--triplet splittings from DFT and CASPT2 are shown in Table \ref{splittings}. The triplet is found to be 21 kJ/mol more stable than the singlet state with CAS(14,16)PT2, independent of the basis set. On the other hand, the DFT calculations predict that the singlet and triplet are essentially degenerate: TPSS, TPSSh and B3LYP predict the singlet to be 4--7 kJ/mol more stable than the triplet, whereas M06-L predicts the triplet to be 1 kJ/mol more stable than the triplet. We note that the splitting was also predicted to be quite small with the larger model 2 and enlarging the model stabilizes the triplet by 8--14 kJ/mol,  making the triplet slightly more stable in all cases (cf.~Table \ref{splittings_model2}). Hence, we can expect --21 kJ/mol to be an lower limit for the singlet--triplet splitting, which is likely to increase for a larger model, assuming CASPT2 behaves similar to  DFT in this regard. Since the splitting is already low, both singlet and triplet states may participate in the mechanism. We can conclude that the singlet--triplet is indeed small as predicted by DFT, and the DFT methods are in this regard in reasonable agreement with CASPT2 for the oxyl species. 

\subsection{The hydroxyl state}

Finally, we investigated also the \ce{[CuOH]^{2+}} hydroxy species. As described in Ref.~\citenum{hedegaard2018}, the \ce{[CuOH]^{2+}} species is readily formed by protonation of \ce{[CuO]+} and may also participate in the \ce{C-H} abstraction in LPMOs.  The \ce{[CuOH]^{2+}} moiety might consist of \ce{OH-} bound to Cu(III). Alternatively, it could be described as a hydroxyl \ce{OH} radical bound to doublet Cu(II), giving either a triplet or an open-shell singlet. Finally, the species may also be formulated as a singlet \ce{OH+} bound to Cu(I). With DFT, we obtained both the triplet and the singlet, the latter in both closed- and open-shell form. However, both the structure and the energy of the open-shell singlet turned out to be almost identical to those of the closed-shell singlet and the spin densities were small, and we therefore discuss only the closed-shell singlet for the DFT calculations.  

\begin{figure}
\centering
\includegraphics[scale=.8]{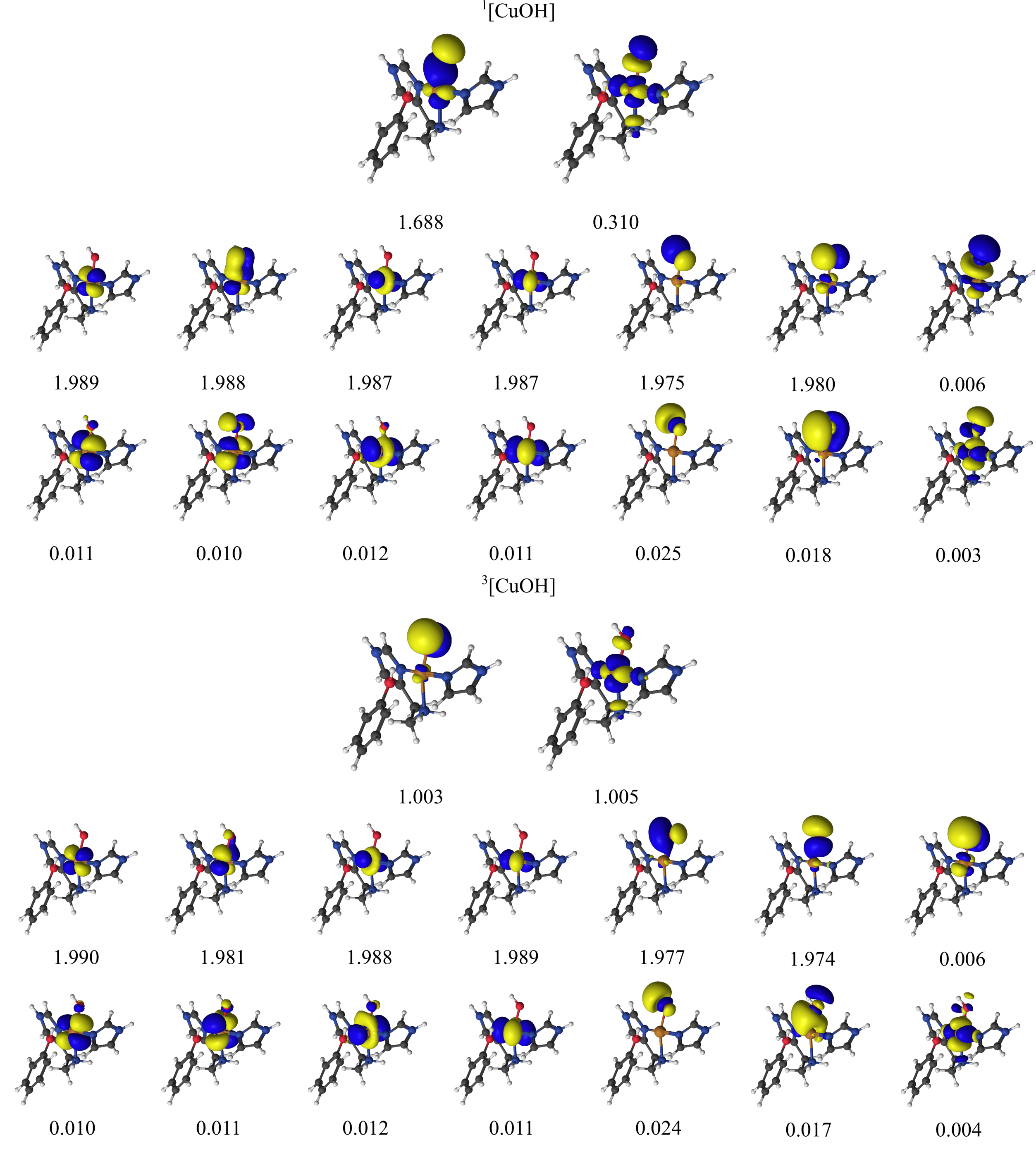}
\caption{Active orbitals and their occupation numbers from a CAS(14,16) calculation with  ANO-RCC type basis sets for the \ce{[CuOH]^{2+}} species. } \label{cu-oh-edl-figure}
\end{figure}

In the CASPT2 calculations, we employed a CAS(14,16) active-space with orbitals and  corresponding natural occupation numbers shown in Figure \ref{cu-oh-edl-figure}. We use the same active space as for the \ce{[CuO]+} species, including five Cu $3d$ orbitals, five correlating Cu $4d$ orbitals, a pair of bonding and anti-bonding orbitals on the \ce{OH} ligand  as well as two lone-pair orbitals with two correlating O $3p$ orbitals.  
The two O $2p$ orbitals and four of the Cu $3d$ orbitals were found to be doubly occupied with occupation numbers close to two for both the singlet and the triplet. The fifth Cu $3d$ orbital ($3d_{z^2}$) and one O $2p$ orbital were singly occupied in the triplet state with occupation numbers of 1.00, indicating a OH radical bound to a Cu(II) coupled to a total triplet state. The singlet state is again somewhat more complicated: the two orbitals having occupation numbers with significant deviation from 2.0 indicates that the singlet is \textit{not} closed-shell. The two partially occupied orbitals (with occupation numbers 1.688 and 0.310, respectively) have large character of both $3d$- and $2p$ atomic orbitals, although to varying degree (see Table \ref{d-orbital-contributions}). Investigating the underlying wave function shows (as expected) that more than one configuration contribute to the ground-state wavefunction: two configurations have large weights of 0.80 and 0.14, respectively. In the configuration with the largest weight, the four $3d$ orbitals as well as the three \ce{OH}-based orbitals with occupation numbers 1.975, 1.980 and 1.688 are double occupied. For the configuration with smaller weight (0.14), the electrons in the orbital with occupation number 1.688 are now interchanged with the orbital with occupation number 0.310, which then becomes doubly occupied. Since the two orbitals with occupation numbers  1.688 and 0.310 are shared between Cu $3d$ and \ce{OH} orbitals, we again suggest that the singlet state has one electron on the Cu center, which is bound to a \ce{OH} radical, spin-coupled to a singlet i.e.~Cu(II) and \ce{OH}. 

The calculated singlet--triplet splittings are given in Table \ref{splittings} and it can be seen that CASPT2 predicts the singlet to be the more stable state by 92--93 kJ/mol. All DFT methods agree with this conclusion, but they overestimate the stability of the triplet significantly. The results depend quite strongly on the DFT functional, but the results shows differences of 31--62 kJ/mol from CASPT2 and are thus significantly off. The TPSS functional gives results closest to CASPT2 (with a difference of 31 kJ/mol), whereas  B3LYP gives the largest difference (62 kJ/mol).

The effect of enlarging the model system was again investigated by DFT and the results are compiled in   Table \ref{splittings_model2}. The larger model gives between 26 kJ/mol (TPSS) and 10 kJ/mol (B3LYP) lower singlet--triplet splitting energies, i.e. always favoring the triplet state. However, even by assuming the largest of these truncation errors (i.e.~the one obtained with the TPSS functional), the singlet is still expected to be significantly more stable than the triplet.    

In our previous study\cite{hedegaard2018}, the reaction in which \ce{[CuOH]^{2+}} abstracts \ce{C-H} from the substrate was found to involve both the singlet and triplet potential-energy surfaces and the reaction energy showed a rather strong dependence on employed the exchange--correlation functional.\cite{hedegaard2018} The present results suggest that part of the reason for this dependency is that the employed DFT functionals struggle to describe the electronic structure of the \ce{[CuOH]^{2+}} singlet correctly. 


\section{Discussion}

In this section, we compare the three intermediates in terms of calculated spin-densities and relate these findings to the findings from the previous section. We also comment on the expected accuracy for CASPT2 for  spin-state splittings of transition metals. 

Analyses of the most important configurations, as well as the active space orbitals for both triplet and singlet spin-states in the \ce{[CuO2]+}, \ce{[CuO]+} and \ce{[CuOH]^{2+}} intermediates, in all cases suggested a Cu(II) oxidation state. This fits well with the Mulliken spin-densities (shown in Table \ref{mulliken_spin}), showing a large spin population on the copper atom for the triplet states ($\sim0.9$ for CAS wave functions and around 0.4--0.6 for DFT). The spin-density obtained from CAS is quite similar for the three intermediates in their triplet spin states (the singlet gives per definition all zero). In addition, Table \ref{mulliken_spin} includes the total Mulliken 3$d$ occupations from the copper atom, which for all intermediates and spin-states are around 9, also in good correspondence with a Cu(II) ion.    

Similar to the spin-densities obtained from CASSCF wavefunctions, the DFT calculations show large spin-density on both copper and the oxygen ligands. Together with the total d-occupations around 9 to 9.5, this is in accordance with a Cu(II) interpretation. However, the variation among the three intermediates is larger for DFT than for CASSCF. 
We note that the open-shell singlet spin-states ($^1$\ce{[CuO2]+} and $^1$\ce{[CuO]+}) calculated with DFT broken-symmetry calculations -- contrary to a CAS wave function -- give rise to a (non-physical) spin-density. While this is a known artifact from the broken-symmetry approach, the resulting spin densities are nevertheless reported in Table \ref{mulliken_spin} to confirm that we obtained a broken-symmetry state. 
\begin{table}[htb!]
\centering
\caption{Mulliken spin-densities employing either a CAS wavefunction (with ANO basis set) or DFT (with def2-TZVPP basis set). For Cu, the total d-occupations are given in parentheses. The atom labels are shown in Figure \ref{structure}. \label{mulliken_spin}} 
\begin{tabular}{lccc|cccc}
\hline\hline
\\[-2.0ex]
Method & \multicolumn{3}{c}{CASSCF}    & \multicolumn{3}{c}{DFT (TPSS)} \\        
\hline \\[-2.0ex]  
State	                     & Cu & O$_1$ &  O$_2$ &  Cu & O$_1$ & O$_2$  \\[1.0ex]             
\hline \\[-1.5ex]
$^3$\ce{[CuO2]+}     & 0.86 (9.05) & 0.25 & 0.80  &  0.35 (9.67)  & 0.71  & 0.77  \\[0.5ex] 
$^1$\ce{[CuO2]+}     & 0.00 (9.05) & 0.00 & 0.00  & -0.33 (9.66)  & 0.13  & 0.35  \\[0.5ex] 
$^3$\ce{[CuO]+}      & 0.84 (9.02) & 1.10 & -     &  0.54 (9.57)  & 1.23  & -     \\[0.5ex] 
$^1$\ce{[CuO]+}      & 0.00 (9.06) & 0.00 & -     & -0.33 (9.56)  & 0.46  & -     \\[0.5ex] 
$^3$\ce{[CuOH]^{2+}} & 0.92 (8.95) & 1.02 & -     &  0.58 (9.53)  & 0.54  & -      \\[0.5ex]
$^1$\ce{[CuOH]^{2+}} & 0.00 (8.99) & 0.00 & -     &  0.00 (9.44)  & 0.00  & -  \\[0.5ex]
\hline \hline
\end{tabular}
\end{table}

The TPSS, TPSSh and B3LYP methods show the expected trend of increasing stability of the triplet state with the increasing amount of Hartree--Fock exchange in the methods (0, 10 and 20\%) for all three complexes, but the heavily parametrized M06-L method (also 0\%) does not follow this trend.
On average, TPSS gives the most accurate results, with an mean absolute error of 26 kJ/mol, but the other methods are not much worse (32--36 kJ/mol). TPSS also has the smallest maximum error (31 kJ/mol), whereas B3LYP gives the largest error (63 kJ/mol).

 Although CASPT2 is known to be a highly accurate quantum chemical model, several studies have shown that it occasionally overestimates the stability of high-spin states.\cite{pierloot2003,vancoillie2010,daku2012} This was recently ascribed to the fact that CASPT2 can be less accurate for the correlation originating from  the metal $3s3p$ orbitals.\cite{pierloot2017} This would imply that for \ce{[CuO2]+} and \ce{[CuOH]^{2+}} (for which the singlet was most stable with CASPT2), the singlet may be even lower than obtained here, while the splitting for \ce{[CuO]+} (where the triplet was most stable) could be smaller than calculated here. The effect was estimated to 8--12 kJ/mol in the benchmark study by Pierloot et al.\cite{pierloot2017}, but it should be noted that the employed benchmarks set did not include any LPMO active-site model (or any other copper system).  It should further be noted that a Dunning type basis set was recommended in Ref.~\citenum{pierloot2017}, prompting us to  apply both ANO and Dunning-type basis sets. However, the obtained differences between the two types of basis sets were very small for our LPMOs models (less than 1 kJ/mol).  
  
A remedy for inaccuracies in the $3s3p$ correlation with CASPT2 was recently suggested in which the $3s3p$ correlation was obtained employing a coupled cluster wavefunction\cite{phung2018}. Yet, due to the multiconfigurational nature of the complexes in this paper, we have refrained from this procedure.

\section{Conclusion}

In this work, we have investigated three intermediates of the LPMO enzyme with CASPT2, namely \ce{[CuO2]+}, \ce{[CuO]+} and \ce{[CuOH]^{2+}} with structures taken from previous QM/MM optimizations.\cite{hedegaard2018}  The first species is expected to be a precursor for generation of the two later intermediates, which probably are involved in the abstraction of a hydrogen atom from the polysaccharide substrate. 

To obtain a first indication of the accuracy of DFT, we have calculated spin-state energy splittings of the three species, which can attain both singlet and triplet spin states;  both spin states are suspected to be involved in the catalytic turnover.
 
Comparing the obtained energies (cf.~Table \ref{splittings}), we see that for both \ce{[CuO2]+} and \ce{[CuO]+}, TPSS, TPSSh and B3LYP functionals give rise to some errors, altough they can still be considered reasonable (errors of 18--33 kJ/mol). An analysis of the underlying  CASSCF wavefunctions shows that the systems are multiconfigurational. The M06-L functional overestimates the stability of the triplet state of\ce{[CuO2]+} by 33 kJ/mol, but is more accurate for \ce{[CuO]+} (19 kJ/mol error). In the case of \ce{[CuOH]^{2+}}, DFT significantly underestimates the singlet--triplet splitting (by 31--61 kJ/mol), and this intermediate is also found to be multiconfigurational.  

Thus we can conclude that the DFT methods for LPMO intermediates can give significant errors (the range found here is 18--62 kJ/mol), reflecting the multiconfigurational nature of the studied intermediates. Perhaps an even worse finding is that DFT sometimes overestimates the stability of the singlet state, while in other cases it is the stability of the triplet state that is overestimated: for instance, all DFT methods overestimate the stability of the triplet state for \ce{[CuO2]+}, but overestimate the stability of the singlet state for \ce{[CuO]+}. Further, all DFT methods significantly underestimate the singlet--triplet splitting for \ce{[CuOH]^{2+}} and produce a qualitatively wrong (closed-shell) description of the singlet state.

These results show that care must be taken when employing DFT for LPMOs. In future studies, we aim at employing more accurate methods also for reactions involving \ce{[CuOH]^{2+}} in abstraction of hydrogen from the polysaccharide substrate.

\begin{acknowledgement}

The authors acknowledge grants from the Swedish research council (project 2018-05003), The Carlsberg Foundation (CF15-0208 and CF16-0482), the European Commission (MetEmbed project 745967), the Royal Physiographic Society in Lund, the China Scholarship Council, and COST through Action CM1305 (ECOSTBio). The computations were performed on computer resources provided by the Swedish National Infrastructure for Computing (SNIC) at Lunarc at Lund University and HPC2N at Ume\aa\  University.

\end{acknowledgement}


\section{Supporting Information Available}

The supporting information contains Mulliken charges as well as XYZ-files for model 1 of the three investigated intermediates. 


\bibliography{lpmo}

\newpage

\begin{center}
\Huge Supporting Information
\end{center}

\begin{table}[htb!]
\centering
\caption{Mulliken charges obtained with ANO or Dunning basis, employing a CAS wavefunction. } 
\label{mulliken_charges} 
\begin{tabular}{lccccc|cccccc}
\hline\hline
\\[-2.0ex]     
&  \multicolumn{5}{c}{ANO} &  \multicolumn{5}{c}{Dunning}      \\[0.5ex] 
\hline \\[-2.0ex]   
                     & Cu & O$_1$ &  O$_2$ & N$_3$ & O$_{Tyr}$ &  Cu   & O$_1$ & O$_2$ & N$_3$ & O$_{Tyr}$  \\[1.0ex]             
\hline \\[-1.5ex]
$^3$\ce{[CuO2]+}     & 0.85 & -0.49 & -0.22 & -0.23 & -0.37    & 1.06 & -0.49 & -0.20 & -0.29 & -0.37    \\[0.5ex] 
$^1$\ce{[CuO2]+}     & 0.83 & -0.49 & -0.20 & -0.23 & -0.37    & 1.04 & -0.49 & -0.19 & -0.29 & -0.37    \\[0.5ex] 
$^3$\ce{[CuO]+}      & 0.83 & -0.60 & -     & -0.23 & -0.38    & 0.94 & -0.53 & -     & -0.30 & -0.35    \\[0.5ex] 
$^1$\ce{[CuO]+}      & 0.84 & -0.65 & -     & -0.23 & -0.38    & 1.03 & -0.56 & -     & -0.30 & -0.35    \\[0.5ex] 
$^3$\ce{[CuOH]^{2+}} & 0.90 & -0.08 & -     & -0.25 & -0.41    & 1.08 & -0.12 & -     & -0.32 & -0.39    \\[0.5ex]
$^1$\ce{[CuOH]^{2+}} & 0.85 & -0.22 & -     & -0.24 & -0.40    & 1.03 & -0.24 & -     & -0.30 & -0.39    \\[0.5ex]
\hline \hline
\end{tabular}
\end{table}

\begin{table}[htb!]
\centering
\caption{Mulliken charges for the three states in triplet and singlet spin-states, obtained with DFT (B3LYP or TPSS). All reported charges are obtained with def2-TZVPP basis set.} 
\label{mulliken_charges_dft} 
\begin{tabular}{lccccc|ccccc}
\hline\hline
\\[-2.0ex]
 & \multicolumn{5}{c}{TPSS}    & \multicolumn{5}{c}{B3LYP} \\        
\hline \\[-2.0ex]  
State                & Cu    & O$_1$   &  O$_2$ & N$_3$  & O$_{Tyr}$ & Cu    & O$_1$  & O$_2$  & N$_3$   & O$_{Tyr}$  \\[1.0ex]             
\hline \\[-1.5ex]
$^3$\ce{[CuO2]+}     & 0.17  &  -0.15  & -0.16  & -0.28  & -0.32     &  0.26 & -0.17  & -0.18  &  -0.28  & -0.32      \\[0.5ex] 
$^1$\ce{[CuO2]+}     & 0.17  &  -0.17  & -0.17  & -0.28  & -0.32     &  0.27 & -0.21  & -0.20  &  -0.27  & -0.35      \\[0.5ex] 
$^3$\ce{[CuO]+}      & 0.15  &  -0.44  & -      & -0.29  & -0.32     &  0.22 & -0.46  & -      & -0.28   & -0.35      \\[0.5ex]
$^1$\ce{[CuO]+}      & 0.15  &  -0.50  & -      & -0.28  & -0.31     &  0.25 & -0.49  & -      & -0.27   & -0.35      \\[0.5ex]
$^3$\ce{[CuOH]^{2+}} & 0.22  &  -0.43  & -      & -0.28  & -0.31     &  0.27 & -0.41  & -      & -0.25   & -0.33      \\[0.5ex] 
$^1$\ce{[CuOH]^{2+}} & 0.13  &  -0.44  & -      & -0.19  & -0.32     &  0.18 & -0.43  & -      & -0.16   & -0.37      \\[0.5ex] 
\hline \hline
\end{tabular}
\end{table}

\newpage

\section*{Coordinates}

\begin{verbatim}
42
[CuO2]+ triplet
N         -5.57900       -4.59700       17.92800
H         -5.83000       -5.59600       17.91500
H         -4.69900       -4.50200       18.47500
C         -5.29800       -4.15100       16.53600
H         -6.18500       -4.37900       15.92400
C         -5.07800       -2.62300       16.52900
H         -4.40300       -2.34200       17.35900
H         -4.56600       -2.31900       15.60800
C         -6.37900       -1.88900       16.65000
N         -7.21500       -2.12700       17.72900
C         -8.29600       -1.36500       17.58500
H         -9.15100       -1.34900       18.25800
N         -8.19600       -0.62700       16.45200
H         -8.87660        0.05010       16.13830
C         -6.99300       -0.95400       15.84100
H         -6.68900       -0.48600       14.90800
H         -5.63510       -7.02140       22.68370
C         -5.85500       -6.04800       22.22100
N         -5.71800       -4.80700       22.80700
H         -5.42300       -4.59900       23.76800
C         -6.04300       -3.85000       21.90500
H         -6.02900       -2.78700       22.13400
N         -6.38700       -4.41900       20.75300
C         -6.28400       -5.78300       20.93400
H         -6.53000       -6.48900       20.13900
H         -8.04140       -9.38100       15.61130
C         -8.16100       -8.47600       16.22500
C         -8.05900       -7.19600       15.64000
H         -7.85300       -7.09400       14.56800
C         -8.24200       -6.03700       16.40000
H         -8.20800       -5.04700       15.94300
C         -8.51500       -6.12500       17.77300
O         -8.64300       -4.93400       18.45200
H         -9.24000       -4.99300       19.27000
C         -8.63300       -7.38800       18.37600
H         -8.87500       -7.45500       19.44100
C         -8.45700       -8.54700       17.60100
H         -8.55800       -9.52900       18.08300
Cu        -6.96700       -3.51500       19.08400
O         -8.44500       -2.61900       20.23400
O         -8.05200       -2.04500       21.30500
H         -4.42090       -4.66390       16.14140
\end{verbatim}

\begin{verbatim}
42
[CuO2]+ singlet
N         -5.58100       -4.58600       17.93600
H         -5.84200       -5.58300       17.93000
H         -4.70000       -4.49500       18.48100
C         -5.29900       -4.15000       16.54000
H         -6.18700       -4.37900       15.93100
C         -5.07600       -2.62300       16.52500
H         -4.40100       -2.33800       17.35400
H         -4.56400       -2.32300       15.60300
C         -6.37700       -1.88900       16.64400
N         -7.21100       -2.13000       17.72400
C         -8.29300       -1.36900       17.58500
H         -9.14400       -1.35500       18.26300
N         -8.19600       -0.62900       16.45400
H         -8.87760        0.04770       16.14160
C         -6.99400       -0.95300       15.83800
H         -6.69100       -0.48100       14.90800
H         -5.63390       -7.02130       22.68330
C         -5.85300       -6.04800       22.22000
N         -5.72000       -4.80500       22.80500
H         -5.42800       -4.59400       23.76700
C         -6.04600       -3.85100       21.90000
H         -6.03900       -2.78600       22.12700
N         -6.38500       -4.42300       20.75000
C         -6.27800       -5.78600       20.93100
H         -6.52500       -6.49400       20.13800
H         -8.04220       -9.37970       15.61170
C         -8.16100       -8.47500       16.22600
C         -8.06000       -7.19500       15.64200
H         -7.85500       -7.09300       14.57000
C         -8.24100       -6.03600       16.40200
H         -8.20800       -5.04600       15.94500
C         -8.51300       -6.12400       17.77600
O         -8.63700       -4.93400       18.45600
H         -9.23600       -4.98700       19.27500
C         -8.63300       -7.38800       18.37800
H         -8.87400       -7.45600       19.44300
C         -8.45700       -8.54700       17.60200
H         -8.55900       -9.52800       18.08400
Cu        -6.96600       -3.50600       19.09100
O         -8.43600       -2.63700       20.23100
O         -8.05600       -2.06600       21.31800
H         -4.42350       -4.66750       16.14800
\end{verbatim}

\begin{verbatim}
41
[CuO]+ triplet
N         -5.61700       -4.52900       17.97100
H         -5.92100       -5.51400       17.99000
H         -4.72500       -4.46500       18.50300
C         -5.32900       -4.13900       16.56100
H         -6.21700       -4.37600       15.95800
C         -5.09600       -2.61300       16.51100
H         -4.41500       -2.31800       17.33200
H         -4.58800       -2.33000       15.58300
C         -6.39600       -1.87500       16.62500
N         -7.23300       -2.11600       17.70000
C         -8.31200       -1.35200       17.57000
H         -9.15800       -1.33900       18.25400
N         -8.21400       -0.61000       16.44000
H         -8.89620        0.06740       16.13030
C         -7.01200       -0.93500       15.82300
H         -6.70900       -0.46300       14.89300
H         -5.63670       -7.01450       22.68890
C         -5.86600       -6.04100       22.23100
N         -5.78000       -4.80000       22.83000
H         -5.50000       -4.59100       23.79600
C         -6.11400       -3.84700       21.92700
H         -6.13700       -2.78000       22.14400
N         -6.40400       -4.42400       20.76300
C         -6.27000       -5.78300       20.93500
H         -6.47700       -6.49000       20.13000
H         -8.04580       -9.37130       15.61630
C         -8.16600       -8.46500       16.22800
C         -8.06900       -7.18500       15.64500
H         -7.87000       -7.08300       14.57200
C         -8.24400       -6.02500       16.40500
H         -8.21600       -5.03600       15.94600
C         -8.50600       -6.10800       17.78200
O         -8.61900       -4.91800       18.45900
H         -9.21200       -4.97400       19.28000
C         -8.62300       -7.37300       18.38300
H         -8.85600       -7.44200       19.45000
C         -8.45400       -8.53400       17.60700
H         -8.55200       -9.51500       18.09100
Cu        -6.93000       -3.36700       19.16300
O         -7.82600       -2.29500       20.43700
H         -4.45620       -4.67320       16.18550
\end{verbatim}

\begin{verbatim}
41
[CuO]+ singlet
N         -5.62100       -4.52700       17.98400
H         -5.93800       -5.50700       18.00700
H         -4.72500       -4.47000       18.51000
C         -5.33500       -4.14600       16.57000
H         -6.22700       -4.38200       15.97100
C         -5.09500       -2.62300       16.51300
H         -4.40700       -2.32600       17.32800
H         -4.59500       -2.34700       15.57900
C         -6.39400       -1.88900       16.63800
N         -7.21400       -2.13400       17.72400
C         -8.30000       -1.37900       17.61100
H         -9.13200       -1.37500       18.31200
N         -8.22300       -0.63800       16.47800
H         -8.91130        0.03380       16.16980
C         -7.02700       -0.95500       15.84300
H         -6.73800       -0.47800       14.91000
H         -5.64810       -7.00000       22.68450
C         -5.88000       -6.02500       22.23100
N         -5.79500       -4.78700       22.83600
H         -5.51600       -4.58200       23.80300
C         -6.12500       -3.82600       21.94200
H         -6.15400       -2.76200       22.16500
N         -6.40500       -4.39700       20.77200
C         -6.27700       -5.76000       20.93400
H         -6.48000       -6.46200       20.12500
H         -8.04770       -9.37110       15.61940
C         -8.17000       -8.46500       16.23100
C         -8.08200       -7.18600       15.64300
H         -7.89200       -7.08500       14.56800
C         -8.25700       -6.02500       16.40200
H         -8.23600       -5.03700       15.93900
C         -8.50800       -6.10500       17.78100
O         -8.61400       -4.91400       18.45800
H         -9.20800       -4.96300       19.27900
C         -8.61800       -7.37000       18.38500
H         -8.84600       -7.43500       19.45400
C         -8.45000       -8.53100       17.61100
H         -8.54300       -9.51200       18.09700
Cu        -6.91400       -3.38200       19.19200
O         -7.84700       -2.37500       20.43900
H         -4.46600       -4.68630       16.19460
\end{verbatim}

\begin{verbatim}
42
[CuOH]2+ triplet
N         -5.67000       -4.48600       18.04700
H         -5.97000       -5.46700       18.13400
H         -4.78600       -4.37400       18.58800
C         -5.39600       -4.16500       16.62000
H         -6.28900       -4.43100       16.03400
C         -5.16200       -2.64100       16.51300
H         -4.48000       -2.31600       17.32400
H         -4.65500       -2.38800       15.57500
C         -6.46300       -1.90000       16.59600
N         -7.36000       -2.15600       17.61800
C         -8.40500       -1.34400       17.46000
H         -9.30400       -1.31800       18.07400
N         -8.23100       -0.57500       16.36200
H         -8.88220        0.13140       16.05050
C         -7.01500       -0.91500       15.79600
H         -6.65300       -0.42500       14.89700
H         -5.66720       -6.97570       22.67500
C         -5.92700       -6.00800       22.22100
N         -5.81900       -4.75400       22.79600
H         -5.51200       -4.51900       23.75000
C         -6.22000       -3.81600       21.90900
H         -6.27800       -2.75400       22.13900
N         -6.57000       -4.40600       20.76600
C         -6.40300       -5.76600       20.94500
H         -6.63900       -6.48900       20.16300
H         -8.04420       -9.37170       15.61110
C         -8.16300       -8.46200       16.21800
C         -8.10500       -7.18300       15.62100
H         -7.95100       -7.08600       14.54000
C         -8.27900       -6.02100       16.37800
H         -8.29700       -5.03800       15.90500
C         -8.50100       -6.10300       17.76200
O         -8.64000       -4.91200       18.44600
H         -9.22700       -4.99700       19.27100
C         -8.58300       -7.36500       18.37600
H         -8.79900       -7.42900       19.44700
C         -8.41500       -8.52600       17.60500
H         -8.49300       -9.50600       18.09400
Cu        -7.08400       -3.33700       19.15400
O         -7.84200       -2.07900       20.40300
H         -8.49400       -1.38000       20.12900
H         -4.52450       -4.71370       16.26300
\end{verbatim}

\begin{verbatim}
42
[CuOH]2+ singlet
N         -5.67900       -4.48300       18.01800
H         -6.05700       -5.44200       18.07800
H         -4.79100       -4.42800       18.56300
C         -5.39600       -4.13200       16.60000
H         -6.28500       -4.38000       16.00400
C         -5.13800       -2.61400       16.53300
H         -4.43600       -2.31300       17.33400
H         -4.65800       -2.34200       15.58600
C         -6.44100       -1.88900       16.66100
N         -7.31700       -2.16300       17.70200
C         -8.40400       -1.39600       17.55500
H         -9.30200       -1.42000       18.17000
N         -8.26500       -0.63300       16.45200
H         -8.93780        0.04940       16.13310
C         -7.04600       -0.93600       15.86800
H         -6.70800       -0.44000       14.96100
H         -5.66050       -6.94650       22.65400
C         -5.88900       -5.99500       22.19700
N         -5.82400       -4.74000       22.77800
H         -5.56900       -4.50200       23.74700
C         -6.16500       -3.79800       21.87300
H         -6.24400       -2.73700       22.10200
N         -6.42900       -4.39300       20.70800
C         -6.27900       -5.75700       20.88900
H         -6.48100       -6.47800       20.09700
H         -8.04580       -9.35610       15.62320
C         -8.16100       -8.46400       16.22100
C         -8.10900       -7.18500       15.62800
H         -7.96200       -7.08400       14.54700
C         -8.27800       -6.02300       16.39100
H         -8.30100       -5.04000       15.91700
C         -8.48500       -6.10700       17.77700
O         -8.59400       -4.91500       18.47300
H         -9.20800       -4.97700       19.27900
C         -8.55900       -7.37000       18.38600
H         -8.76200       -7.43600       19.46000
C         -8.40000       -8.53100       17.60900
H         -8.47200       -9.51100       18.09800
Cu        -6.96300       -3.35600       19.14300
O         -7.93200       -2.33300       20.35400
H         -8.36200       -1.51400       20.00200
H         -4.52810       -4.68250       16.23700
\end{verbatim}

\end{document}